\documentclass[conference]{IEEEtran}
\IEEEoverridecommandlockouts
\def\BibTeX{{\rm B\kern-.05em{\sc i\kern-.025em b}\kern-.08em
    T\kern-.1667em\lower.7ex\hbox{E}\kern-.125emX}}

\usepackage[linesnumbered,ruled,vlined]{algorithm2e}
\usepackage{multicol}
\usepackage{multirow}
\usepackage{makecell}
\usepackage{url}

\usepackage{cite}
\usepackage{amsmath,amssymb,amsfonts}
\usepackage{algorithmic}
\usepackage{graphicx}
\usepackage{textcomp}
\usepackage{xcolor}

\begin{document}
\title{Efficient Kernelization Algorithm for Bipartite Graph Matching}

\author{Guang Wu$^\dagger$, Xinbiao Gan$^\dagger$, Zhengbin Pang$^\dagger$, Bo Huang$^\dagger$, Bopin Ran$^\dagger$\\
$^\dagger$National University of Defence Technology, College of Computer Science and Technology\\
wuguang@nudt.edu.cn}

\maketitle
\begin{abstract}
Finding the maximum matching in bipartite graphs is a fundamental graph operation widely used in various fields. To expedite the acquisition of the maximum matching, Karp and Sipser introduced two data reduction rules aimed at decreasing the input size. However, the KaSi algorithm, which implements the two data reduction rules, has several drawbacks: a high upper bound on time complexity and inefficient storage structure. The poor upper bound on time complexity makes the algorithm lack robustness when dealing with extreme cases, and the inefficient storage structure struggles to balance vertex merging and neighborhood traversal operations, leading to poor performance on real-life graphs. 

To address these issues, we introduced MVM, an algorithm incorporating three novel optimization strategies to implement the data reduction rules. Our theoretical analysis proves that the MVM algorithm, even when using data structures with the worst search efficiency, can still maintain near-linear time complexity, ensuring the algorithm's robustness. Additionally, we designed an innovative storage format that supports efficient vertex merging operations while preserving the locality of edge sets, thus ensuring the efficiency of neighborhood traversals in graph algorithms. 

Finally, we conduct evaluations on both real-life and synthetic graphs. Extensive experiments demonstrate the superiority of our method.
\end{abstract}

\section{Introduction}

As a fundamental task in graph analysis, finding the maximum matching in a bipartite graph has wide applications in various fields, such as complex network analysis~\cite{Liu2011ControllabilityOC,Zhu2024MinimalCN,Vazifeh2018AddressingTM}, resource allocation~\cite{Harder2023StrategicRS,Bachor2023TheMT}, subgraph matching~\cite{Jiang2024,Choi2023BICEEC,Rivero2017EfficientAS}, sparse linear systems~\cite{Commault2024DilationCS,Davis2016ASO,Pothen1990ComputingTB} and crowdsourcing\cite{Ren2023,Lai2022LoyaltybasedTA,Cheng2020,Tong2017}. However, with the continuous expansion of graph data scales, directly applying exact algorithms to find the maximum matching faces numerous challenges~\cite{Duff2011DesignIA,Kaya2020KarpSipserBK,Panagiotas2020EngineeringFA}. Consequently, researchers have extensively focused on accelerating the acquisition of the maximum matching through initialization methods~\cite{Abu-Khzam2022,Pothen2019ApproximationAI}. Among these methods, the data reduction rules proposed by Karp and Sipser are the most renowned~\cite{Himmel2024FastPP,Kaya2020KarpSipserBK,Karp1981MaximumMI,Abu-Khzam2022}. These rules transform the original problem input into a kernel graph using vertex removal and merging techniques, thereby reducing the time consumption of subsequent exact algorithms~\cite{Abu-Khzam2022}.

Vertex removal operations are straightforward, but vertex merging operations present certain challenges~\cite{Himmel2024FastPP,Abu-Khzam2022}. In vertex merging, in addition to merging the edge tables of the two vertices being processed, the connectivity status of the vertices in these edge tables must also be updated, resulting in significant time consumption. Repeatedly merging the same vertices can lead to a time complexity upper bound of $O(n^2)$, which is too high for sparse graphs (where $m << n^2$)\cite{brandstadt1999graph} and may cause a lack of robustness when handling extreme instances. Kaya et al.\cite{Kaya2020KarpSipserBK} proposed two theoretical algorithms, $HKaSi$ and $TKaSi$, which utilize hash tables or binary search trees as the storage format for edge tables. These algorithms achieve time complexity upper bounds of $O(m \log n)$ and $O(m \log^2 n)$, respectively. However, these storage structures are not well-suited for graph computation, as graph algorithms often involve substantial neighborhood traversals (sequential access)\cite{chhugani2012fast}, while these structures lack data locality. In fact, all current implementations of the $KaSi$ algorithm, similar to the current graph computation\cite{fan2021graphscope,xinbiaogan24,Islam2023DGAPED,Wheatman2021APP,Leo2019PackedMA} and graph mining systems\cite{graphfold2024,cyclosa2023}, predominantly use storage formats like CSR (Compressed Sparse Row). As a result, the time complexity of these advanced $KaSi$ algorithm variants remains $O(n^2)$, regardless of the optimization strategies they employ.

To ensure the efficiency of traversal operations, we proposed an algorithm, $MVM$, which can maintain near-linear time complexity even on data structures with the worst search efficiency, to implement the data reduction rules of Karp and Sipser. We introduced three optimization strategies: reducing the number of graph structure modifications through a multi-vertex merging strategy, minimizing irrelevant overhead during the search for mergeable vertices using indirect set operations, and avoiding excessive time consumption caused by repeatedly processing high-degree vertices with a balanced processing strategy. We then theoretically proved that $MVM$, incorporating these three strategies, can  maintain $O(min(m\log n,n^2))$, even with storage structures having $O(n)$ search efficiency ($HKaSi$ and $TKaSi$ achieve better time complexity because hash tables and binary search trees have $O(1)$ and $O(logn)$\cite{west2001introduction} search efficiencies, respectively). In addition, we designed an efficient data structure to support vertex merging operations. This data structure is implemented based on the CSR format, using connecting edge tables and batch updates to perform vertex merging, which effectively avoids frequent memory allocations and data movements. The CSR-like storage format also ensures data locality.

We conducted evaluations on both real-life and synthetic graphs. First, we compared $MVM$ with the state-of-the-art variants of $KaSi$, including $KaSi\_cache$\cite{Kaya2020KarpSipserBK} and $KaSi\_comp$\cite{Koana2021}, as well as the theoretical algorithms $HKaSi$ and $TKaSi$. The experiments demonstrated that our algorithm outperforms all others in all tested instances. On 12 real-life graphs, $MVM$ achieved an average speedup of 25x and 56x over $HKaSi$ and $TKaSi$, respectively. In instances that exhibit the typical time complexity of the $KaSi$ algorithm, $MVM$ achieved a speedup of approximately three orders of magnitude over $KaSi\_cache$ and $KaSi\_comp$. Next, we compared $MVM$ with maximal matching algorithms~\cite{GuangWu24,Panagiotas2020EngineeringFA,Pothen2019ApproximationAI} that also aim to accelerate the acquisition of maximum bipartite matchings. On all real-life graphs~\cite{Davis2011TheUO,snapnets}, $MVM$ consistently achieved more stable speedup. The stability arises because, when using maximal matching algorithms as an initialization method, the subsequent use of exact algorithms may still require traversing the entire graph, leading to less consistent speedups. In contrast, kernelization algorithms reduce the search space, thereby reliably lowering runtime. Extensive experiments demonstrated the superiority of our proposed methods.


In summary, we make the following contributions:

1) We proposed an efficient algorithm, $MVM$, which combines three novel optimization strategies to implement the Karp-Sipser data reduction rules on bipartite graphs. We have demonstrated that its time complexity is $O(min(m \log n, n^2))$. To the best of our knowledge, this is the first algorithm for implementing the Karp-Sipser data reduction rules that can guarantee near-linear time complexity, even when using a data structure with the poorest search efficiency. (Sec. \ref{sec:phe}, \ref{sec:alg})

2) We designed an efficient data structure to support the data reduction rules. This data structure enables efficient vertex merging operations and neighborhood traversal, which are essential for graph computation. (Sec. \ref{sec:data})

3) We compared our approach with other state-of-the-art methods for accelerating the acquisition of the maximum matching on both real-life and synthetic graphs. Extensive experimental results demonstrated the superiority of our proposed algorithm. (Sec.~\ref {sec:exp})

\section{PRELIMINARIES}

In this section, we will introduce the preliminary knowledge relevant to the research problem of this paper. The notation and their definitions used in the paper are listed in Table \ref{tab:nota}.

\noindent\textbf{Problem Definition.}
The task of Bipartite Graph Matching receives a bipartite graph $\mathcal{G}=(\mathcal{V}_C \cup \mathcal{V}_R, \mathcal{E})$ as input, where $\mathcal{V}_C$ and $\mathcal{V}_R$ are two disjoint vertices subsets, and $\mathcal{E} \subseteq \mathcal{V}_C \times \mathcal{V}_R$ is a set of edges. Its output is a matching $M$, where $M \subseteq \mathcal{E}$ and any two edges in $M$ do not depend on the same vertex. A matching $M$ is called maximal if no other matching $M'\supset M$ exists. A maximal matching $M$ is called maximum if $|M| \geq |M'|$ for every matching $M'$. 

 \begin{table}[]
 \renewcommand\arraystretch{1.3}
     \centering
    \caption{Notations}
     \begin{tabular}{c|c}
     \hline  Symbol & Definition \\ \hline
     $\mathcal{G}=(\mathcal{V}_C\cup \mathcal{V}_R, \mathcal{E})$ & The original bipartite graph \\ \hline
     $\mathcal{G}'$ & The reduced/kernelized bipartite graph \\ \hline
$\mathcal{M}$, $\mathcal{M}'$ &  The maximum matching in $\mathcal{G}$ and $\mathcal{G'}$  \\ \hline
n, m & The number of vertices and edges in $\mathcal{G}$ \\ \hline
     
$\Gamma(v)$, $\Gamma(V)$ & The neighboring vertices set of $v$, $\bigcup_{v \in V} \Gamma(v)$\\ \hline
 $|\cdot|$ &  The number of element in set $\cdot$ \\ \hline
 
$deg(v)$ & The degree of vertex $v$\\ \hline
$\hat{V}$, $\Tilde{V}$ & The set of mergeable and boundary vertices \\ \hline
\rule{0pt}{5.5mm}
$m(\Gamma(\hat{V}))$ & \makecell[c]{$\mathcal{G} \setminus \hat{V} * \Gamma(\hat{V})$, Reomve the vertices in $\hat{V}$ \\ and merge the vertices in $\Gamma(\hat{V})$}\\
\hline
     \end{tabular}
     \label{tab:nota}
\end{table}

\noindent\textbf{Karp and Sipser' data reduction rule.} 
\begin{enumerate}
    \item \textbf{Rule 1}: let $deg(u)=1$, then delete $u$ and its neighbor $v$ and add the edge $(u,v)$ to the matching. 
    \item \textbf{Rule 2}: let $deg(u)=2$, then remove $u$, merge its neighbor vertices $v$ and $w$ to $vw$. Karp and Sipser showed that $\mathcal{M}'$ for the reduced graph can be extended to obtain $\mathcal{M}$ for the original graph by matching $u$ with either $v$ or $w$ depending on $vw$' match.   
\end{enumerate}

\begin{figure}
    \centering
    \includegraphics[width=0.9\linewidth]{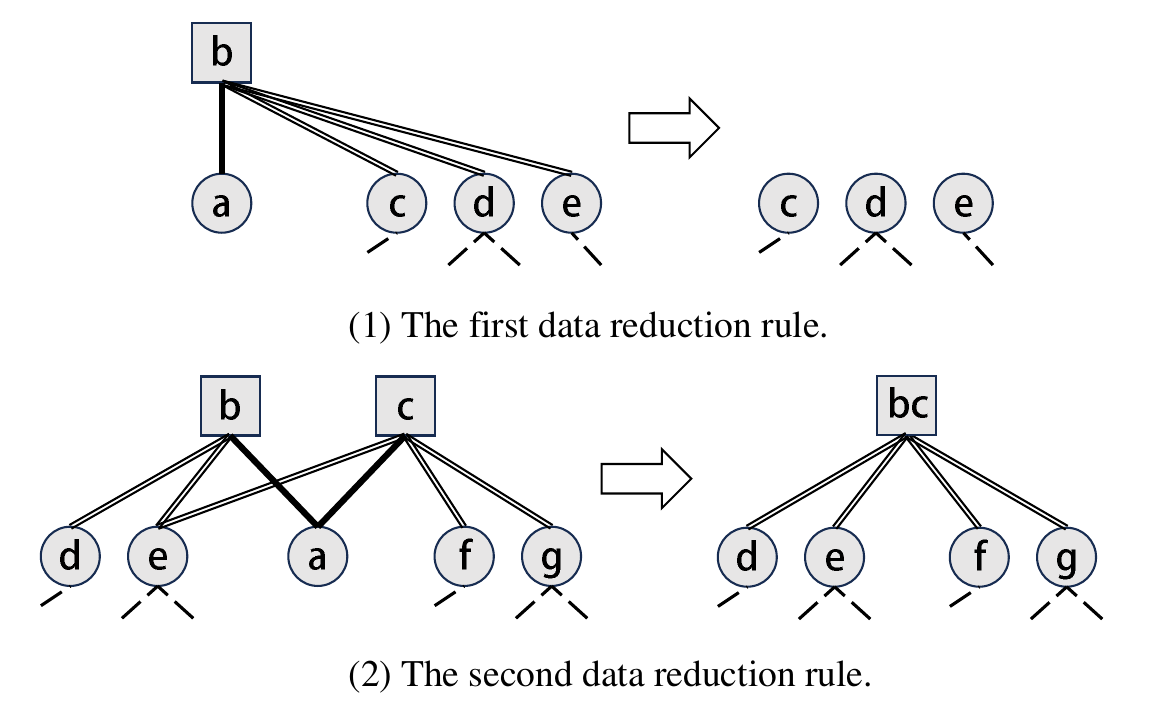}
    \caption{An example of Karp and Sipser's data reduction rule, where the black solid lines represent edges related to the matching, and the black double lines represent edges related to update operations. Circle vertices and square vertices denote the two types of vertices in the bipartite graph.}
    \label{fig:sigmod_kasi_rule}
\end{figure}

We refer to vertices that initially have a degree of 2 in the original graph, or vertices whose degree is reduced to 2 due to data reduction operations, as $\textbf{mergeable vertices}$. The neighboring vertices of these mergeable vertices are called $\textbf{boundary vertices}$. Vertices that are adjacent to boundary vertices but are not mergeable vertices are called $\textbf{external vertices}$. The overall time consumption can be expressed as $T_{KaSi}=\sum_{u \in \mathcal{G}} T(m(\Gamma(u)))$, where $u$ represents mergeable vertices. Obviously, the overall time complexity of the algorithm depends on the number of mergeable vertices $u$ and the cost of each merging operation $m(\Gamma(u))$.

\section{Philosophy}
\label{sec:phe}

In this section, we will introduce the concepts behind our multi-vertex merging strategy and indirect set operation strategy.

\noindent\textbf{Q1, How to reduce the overall time complexity of the KaSi algorithm?}

When we use a storage structure with $O(n)$ search efficiency, the overall time complexity of the KaSi algorithm can be expressed as $T=\sum_{\hat{v} \in \mathcal{G}} \sum_{v \in \Gamma(\hat{v})} deg(v)$, which is the sum of the degrees of the boundary vertices. If the time complexity for processing each boundary vertex cannot be reduced, a natural idea to lower the overall time complexity is to reduce the number of boundary vertices to be processed.


\noindent\textbf{Q2, How to reduce the number of boundary vertices?}

Different mergeable vertices may share the same boundary vertices. Therefore, the number of boundary vertices that need to be handled can be reduced by processing these adjacent mergeable vertices together. The benefits of optimization methods based on components\cite{Langguth2021SharedmemoryIO,Langguth2010HeuristicIF} are essentially due to the reduced number of boundary vertices processed. However, the current component-based method can only identify explicit mergeable vertices, which limits the benefits that such methods can provide. Therefore, we propose a multi-vertex merging strategy to identify all mergeable vertices adjacent to the currently processed mergeable vertices (where adjacency indicates sharing boundary vertices), regardless of whether they are explicit or implicit.

\noindent\textbf{Q3, How to identify implicit mergeable vertices?}

If a vertex is mergeable, then at some stage of the kernelization algorithm, the size of its neighbor set will become two, regardless of the original size. Let $\hat{V}$ be the set of currently identified mergeable vertices. A vertex v is considered to be mergeable if $deg(v)-1$ of its neighbors are already in $\Gamma(\hat{V})$. This is because, through merging operations on vertices in $\hat{V}$, $v$ will become a vertex with a degree of two. Formally, a vertex $v$ that satisfies $|\Gamma(v)-\Gamma(\hat{V})|=1$ is a mergeable vertex (where $\Gamma(\hat{V})= \bigcup_{\hat{v} \in \hat{V}} \Gamma(\hat{v})$). Our multi-vertex merging method aims to identify all adjacent mergeable vertices that meet this criterion, starting from an explicit mergeable vertex. It is easy to see that the component-based method is a special case of our approach.

\noindent\textbf{Q4, How to identify the mergeable vertices for "free"?}

When identifying mergeable vertices through search operations, it is possible to encounter vertices that are not mergeable. Directly checking if these vertices meet the criteria can lead to additional time consumption. Therefore, we propose an indirect processing strategy to perform the set operations required for vertex identification. In multi-vertex merging operations, the time consumed is related to the degrees of all boundary vertices. Therefore, if we can ensure that the edges accessed during the search are all connected to boundary vertices, the process of identifying mergeable vertices will not incur additional overhead. For each vertex encountered during the search process, we determine its mergeability by comparing its degree with the number of times it has been accessed during the current search operation. This ensures that if the vertex is not mergeable, we do not access edges in its edge set that are unrelated to the current boundary vertices. By using this indirect method, we can efficiently identify mergeable vertices without increasing the time complexity of the subsequent multi-vertex merging operation.

Although multi-vertex merging and indirect set operations can theoretically reduce the time consumption of the kernelization algorithm, greedily implementing these two strategies is still insufficient to achieve a lower time complexity. In the next section, we will introduce how to incorporate a balanced processing strategy into our $MVM$ algorithm and, in Section \ref{sec:ana}, demonstrate why the $MVM$ algorithm, combining these three strategies, can achieve a near-linear time complexity.

\section{Algorithm}
\label{sec:alg}
\subsection{The outline of the algorithm.}

\begin{algorithm}
\caption{The overall framework of obtaining the maximum matching}
\label{alg:overall}
\KwIn{The origin bipartite graph $\mathcal{G}$}
\KwOut{The maximum matching $\mathcal{M}$ of $\mathcal{G}$}
\tcp{Kernelize the original graph and record the matching information on the matching tree.}
$\mathcal{G}'$, $\mathcal{T}$ $\gets$ $\texttt{MVM}(\mathcal{G})$\;
\tcp{Find the maximum matching $\mathcal{M}'$ in the kernelized graph $\mathcal{G}'$.}
$\mathcal{M'} \gets \texttt{MM}(\mathcal{G}')$\;
\tcp{Reconstruct the maximum matching $\mathcal{M}$ of the original graph $\mathcal{G}$ based on $\mathcal{M}'$.}
$\mathcal{M} \gets$  $\texttt{RTM}$ ($\mathcal{M'},\mathcal{T}$)\;
\textbf{Return} $\mathcal{M}$\;
\end{algorithm}

In this section, we will outline the overall process of using the kernelization algorithm to accelerate the acquisition of the final maximum matching. As shown in Algorithm \ref{alg:overall}, we first reduce the original graph $\mathcal{G}$ to a kernel graph $\mathcal{G}'$ by invoking the kernelization algorithm $MVM$. During the execution of $MVM$, we record the key edges related to the merging operations in a matching tree $\mathcal{T}$, which will facilitate the reconstruction of the maximum matching of the original graph. Next, we employ an exact algorithm to obtain the maximum matching $\mathcal{M}'$ on the kernel graph $\mathcal{G}'$.  Finally, we reconstruct the maximum matching $\mathcal{M}$ of the original graph based on $\mathcal{M'}$ and $\mathcal{T}$. The algorithm process concludes with the matching $\mathcal{M}$ return.


\subsection{Balanced multi-vertex merging algorithm.}
\begin{algorithm}
\caption{Multi-vertex Merging algorithm}
\label{alg:DBMVM}
\SetKwProg{Fn}{Function}{}{end}
\SetKwFunction{FMains}{MVM}
\Fn{\FMains{$\mathcal{G}$}}{
Initialize the $buckets[1]$, $buckets[2]$ and $buckets[3]$\; 
$round$  $\gets$ 1\;
\For{each $v$ $\in$ $V$}{
$rnd$[$v$] $\gets$ 0\;
}

\While{$buckets[1] \neq \emptyset$ $\vee$ $buckets[2] \neq \emptyset$ $\vee$ $buckets[3] \neq \emptyset$}{
\While{$bucket[1] \neq \emptyset$ }{
$u$ $\gets$ $buckets[1]$, $v$ $\gets$ $\Gamma(u)$\;
$\mathcal{M}$ $\gets$ $\mathcal{M} \cup orig(u, v)$, $\mathcal{G}$ $\gets$ $\mathcal{G}$ $\setminus$ $\left\{u, v\right\}$\;
update the buckets\;
}

\eIf{$bucket[2] \neq \emptyset$}{
$u \gets buckets[2]$\;
$\hat{V}$ $\gets$ $\hat{V} \cup u$, $\Tilde{V}$ $\gets$ $\Tilde{V} \cup \Gamma(u)$\;

$\mathcal{T}$ $\gets$ $\mathcal{T} \cup \left\{orig(u, v)|v \in \Gamma(u)  \right\}$\;

\While{$\exists$ an unprocessed $\Tilde{v}$ $\in$ $\Tilde{V}$  $\wedge$ $|\Tilde{V}| \neq |\hat{V}|$}{
\For{each $\hat{v}$ $\in$ $\Gamma(\Tilde{v})$}{
\If{$|\Gamma(\hat{v}) - \Tilde{V}| \leq 1$ $\wedge$ $rnd[\hat{v}] \neq round$}{
$\hat{V}$ $\gets$ $\hat{V} \cup \hat{v}$, $\Tilde{V}$ $\gets$  $\Tilde{V} \cup \Gamma(\hat{v})$\;

\If{$|\Tilde{V}|=|\hat{V}|$}{
break\;
}
$\mathcal{T} \gets \mathcal{T} \cup orig(\hat{v},\Tilde{v})$\;
$\mathcal{T} \gets \mathcal{T} \cup orig(\hat{v},\Gamma(\hat{v}) - \Tilde{V})$\;

}
}
}
$\texttt{Merge}$($\mathcal{G}$, $\Tilde{V}$, $\hat{V}$, $buckets$, $rnd$, $round$)\;
}
{
swap($buckets[2],buckets[3]$)\;
$round$++\;
}
}
$\textbf{Return}$ $\mathcal{G}'$, $\mathcal{T}$
}
\end{algorithm}

In this section, let's discuss how to incorporate balanced processing strategies into our multi-vertex merging algorithm. As shown in Algorithm \ref{alg:DBMVM}, first, we initialize three buckets for storing processable vertices, where $buckets1$ is used to store vertices of degree 1, and both $buckets2$ and $buckets3$ are used to store vertices of degree 2 (line 1). At the beginning of the algorithm $MVM$, $buckets3$ is empty. During subsequent processing, if new processable vertices of degree 2 are generated and these vertices are related to the merge operation in the current round, they will be added to $buckets3$. Then, we initialize the global processing round $round$ and the current processing round $rnd[v]$ for each vertex $v$ (line 3-5). A vertex $v$ is processable in the current round only if $rnd[v]$ is not equal to $round$.

We continuously process vertices from these three buckets until all buckets are empty (line 6). We prioritize processing and matching vertices with a degree of one (lines 7-10), as the vertex removal operation is relatively straightforward. When $buckets[1]$ is empty, we will search for adjacent mergeable vertices to extend the degree-two vertex $u$. We add vertex $u$ to the set of mergeable vertices $\hat{V}$ and its neighboring vertices to the set of boundary vertices $\Tilde{V}$. The original records of edges related to vertex $u$ are added to the matching tree $\mathcal{T}$ (lines 12-14). We then search from vertex $\Tilde{v}$ in the boundary vertices set to identify new mergeable vertices. For an encountered vertex $\hat{v}$, if $|\Gamma(\hat{v})|-1$ of its neighboring vertices are already included in the boundary vertices set, and $\hat{v}$ is a vertex that can be processed in the current round, it will be added to $\hat{V}$. Subsequently, all neighboring vertices of $\hat{v}$ will be incorporated into $\Tilde{V}$ (line 18).

Afterward, we assess the cardinality of $\Tilde{V}$ and $\hat{V}$. If they are equal, this implies that all neighboring vertices of $\hat{v}$ already exist in boundary vertices set before processing $\hat{v}$. Merging mergeable vertices in $\hat{V}$ will reduce $\hat{v}$'s degree to one, allowing it to be directly removed, enabling us to exit the search prematurely. This early exit mechanism can reduce the algorithm's overall consumption. If not, the original edges of the related edges are recorded into $\mathcal{T}$ (lines 21-22). After meeting the search requirements or when there are no more unprocessed vertices in the $\Tilde{V}$, we perform a multi-vertex merging operation on the vertices in the mergeable vertices set $\hat{V}$ (line 23). 

If all processable vertices in the current round have been handled, we swap the data in $buckets2$ and $buckets3$, update the global processing round, and move to the next round (lines 25-26). When all buckets are empty, we return the kernel graph $\mathcal{G}'$ obtained after multiple vertex removals and merge operations, along with the matching tree $\mathcal{T}$ that records the essential information.

\subsection{Merging the search graph}

\begin{algorithm}
\caption{Merge the vertices}
\label{alg:3}
\SetKwProg{Fn}{Function}{}{end}
\SetKwFunction{FMains}{Merge}
\Fn{\FMains{$\mathcal{G}$, $\Tilde{V}$, $\hat{V}$, $buckets$, $rnd$, $round$}}{
\If{$|\Tilde{V}| = |\hat{V}|$}{
$\hat{v}$ $\gets$ the last vertex added to $\hat{V}$\;
$\hat{V}$ $\gets$ $\hat{V} \setminus \hat{v}$\;
}
$\Tilde{v}$  $\gets$ $\Tilde{V}$\;
$\Gamma(\Tilde{v})$ $\gets$  $\Gamma(\Tilde{v}) \cup \Gamma(\Tilde{V})$\;
\For{each $v$ $\in$ $\Gamma(\Tilde{v})$}{
$\Gamma(v) \gets \Gamma(v) \cup \Tilde{v}$\;
$rnd[v]$ $\gets$ $round$\;
}
$\Tilde{V}$ $\gets$ $\Tilde{V} \backslash \Tilde{v}$\; 
$\mathcal{G}$ $\gets$ $\mathcal{G}$ $\backslash$ $\left\{\hat{V}, \Tilde{V} \right\}$\;
update the $buckets$\;
}
\end{algorithm}

In this section, we will discuss how to implement multi-vertex merging operations. As shown in Algorithm \ref{alg:3}, we first compare the sizes of the mergeable vertices set $\hat{V}$ and the boundary vertices set $\Tilde{V}$. If they are equal, it indicates that the last vertex added to the mergeable vertices set will become a degree-one vertex due to previous merge operations. Therefore, we remove the last vertex added to the mergeable vertices set (lines 2-4).

Next, we begin the multi-vertex merging operation, which consists of establishing new edge connections and removing vertices. First, we select a vertex $\Tilde{v}$ from the boundary vertices set as the merged vertex (this vertex can be the one with the highest degree in $\Tilde{V}$ to minimize the number of edges needing reconnection). Then, we connect all neighboring vertices of the vertices in $\Tilde{V}$ to $\Tilde{v}$ and update their visitation rounds (lines 6-9). In the actual implementation, while establishing these new edge connections, we also record the original source and target vertices of these edges. Finally, we remove all mergeable vertices and boundary vertices (except $\Tilde{v}$) from the graph and add the newly generated processable vertices to the $buckets$ (lines 10-12).

\subsection{Reconstruct the matching}

Reconstructing the maximum matching on the original graph through RTM is straightforward, which we'll briefly introduce in this section. First, we remove all matched vertices on $\mathcal{T}$, then starting from unmatched leaf vertices, we recursively match the entire matching tree $\mathcal{T}$ to recover the maximum matching $\mathcal{M}'$ on the original graph. Similar to repeatedly removing leaf vertices from the tree, it's easy to understand that this operation can be completed in linear time.

\subsection{Time Complexity}
\label{sec:ana}

In this section, we will analyze the time complexity of the $MVM$ algorithm. It is easy to understand that the overall time consumption of the algorithm can be expressed as follows.
\begin{equation}
\begin{tabular}{c}
$T_{MVM}=O(\sum_{\hat{V} \in \mathcal{G}}T(s(\hat{V}))+ T(m(\Gamma(\hat{V}))))$   
\end{tabular}
\end{equation}
Here, $T(s(\hat{V}))$ represents the time spent searching for the mergeable vertex set $\hat{V}$.

Since we are considering the time consumption on a data structure with the worst-case search efficiency, the time cost of the multi-vertex merging operation can be directly represented by the sum of the degrees of all boundary vertices. As discussed in previous sections, we use an indirect set operation strategy to search for mergeable vertices, during the search process we do not access edges that are unrelated to the boundary vertices. Therefore, $T(s(\hat{V}))=O(T(m(\Gamma(\hat{V}))))$, and the overall time consumption can be further expressed as follows.
\begin{equation}
    \begin{tabular}{l}
    $T_{MVM}=O(\sum_{\hat{V}\in \mathcal{G}} T(m(\Gamma(\hat{G}))))$\\[2mm]
    $\qquad\quad=O(\sum_{\hat{V}\in \mathcal{G}} \sum_{\Tilde{v} \in \Gamma(\hat{V})} deg(\Tilde{v}))$
    \end{tabular}
\end{equation}

According to our balanced processing strategy, in each iteration, the mergeable vertices on the same side of the bipartite graph have non-overlapping boundary vertices. Since the overall time complexity of the algorithm is equivalent to the time required to process the mergeable vertices on one side of the bipartite graph, we can further express the total time complexity as follows.
\begin{equation}
\begin{tabular}{l}
$T_{MVM}=O(\sum_{r=1}^{r=R} \sum_{\hat{V}^i_r \in \mathcal{G}}\sum_{\Tilde{v} \in \Gamma(\hat{V}^i_r)} deg(\Tilde{v}))$ \\[1mm]
$\qquad\qquad=O(\sum_{r=1}^{r=R} m)$\\[1mm]
$\qquad\qquad=O(R*m)$
\end{tabular}
\end{equation}
Here, $\hat{V}_r^i$ represents the $i$th set of mergeable vertices on the same side on the bipartite graph in the $r$th iteration.

From the above, it is clear that the key to analyzing the algorithm's time complexity lies in determining the worst-case number of access rounds.

We can observe that during each multi-vertex merging operation, the set of mergeable vertices must include at least one explicit mergeable vertex as the initial vertex. According to our multi-vertex merging strategy, in each round of processing, all explicit mergeable vertices will either be processed as starting vertices or be incorporated into others. Therefore, except for the first round, the explicit mergeable vertices processed in each subsequent round must be implicit in the previous round. In the $(r-1)$th round of processing, the neighbor vertex set of the starting mergeable vertices in the $r$th round should satisfy the following relationship.
\begin{equation}
\begin{tabular}{c}
$\Gamma(\hat{v})= \Tilde{V}_1 \cup \Tilde{V}_2, \Tilde{V}_1 \cap \Tilde{V}_2 =\emptyset$\\
$|\Tilde{V}_1|=k-x,|\Tilde{V}_2|=x,1<x<k-1$\\
$\Tilde{V}_1 \subseteq \Gamma(\hat{V}_1),\Tilde{V}_2 \subseteq \Gamma(\hat{V}_2)$  
\end{tabular}
\end{equation}
Here, $k$ represents the number of neighboring vertices of $\hat{v}$ in the $(r-1)$th round, while $\hat{V}_1$ and $\hat{V}_2$ denote two sets of mergeable vertices. 

The above relation indicates that in the $r$th round, the two boundary vertices of the explicit mergeable vertex $\hat{v}$ are formed by two sequences of multi-vertex merging operations from the previous round. Otherwise, $\hat{v}$ would be merged into one of these vertex sets and processed together. 

Furthermore, in the $r$th round, any other explicitly present mergeable vertices related to $\hat{V}_1$ and $\hat{V}_2$ can be discovered through a search starting from $\hat{v}$. Thus, we can derive the following relationship regarding the number of multi-vertex merging operations in each round.
\begin{equation}
    \begin{tabular}{c}
        $N(\hat{V}_r) \leq N(\hat{V}_{r-1})/2$ \vspace{1ex}\\
        $\sum_{r=1}^{r=R} N(\hat{V}_r) \leq n$
    \end{tabular}
\end{equation}
Here, $N(\hat{V}_r)$ represents the number of multi-vertex merging operations performed in the $r$th round. Based on the above relationship, we can deduce that the number of processing rounds $R$ in the worst-case scenario is $\log n$. Therefore, for $MVM$, we can establish an upper bound on the time complexity as $O(m \log n)$. Additionally, through amortized analysis, it is evident that the algorithm can still maintain a processing cost of $O(n)$ for each mergeable vertex in the multi-vertex merging operation. Consequently, $MVM$ retains an upper bound of $O(n^2)$ for its time complexity, leading to a final time complexity for $MVM$ of $O(min(m \log n,n^2))$.

\section{data structure}
\label{sec:data}
In this section, we present the implementation of vertex merging operations using our proposed data structure. We divide the vertex merging operation into two parts: connecting the edge tables of boundary vertices and updating the states of external vertices. 

\subsection{Connecting the edge tables of the boundary vertices.}

\begin{figure}
    \centering
\includegraphics[width=\linewidth]{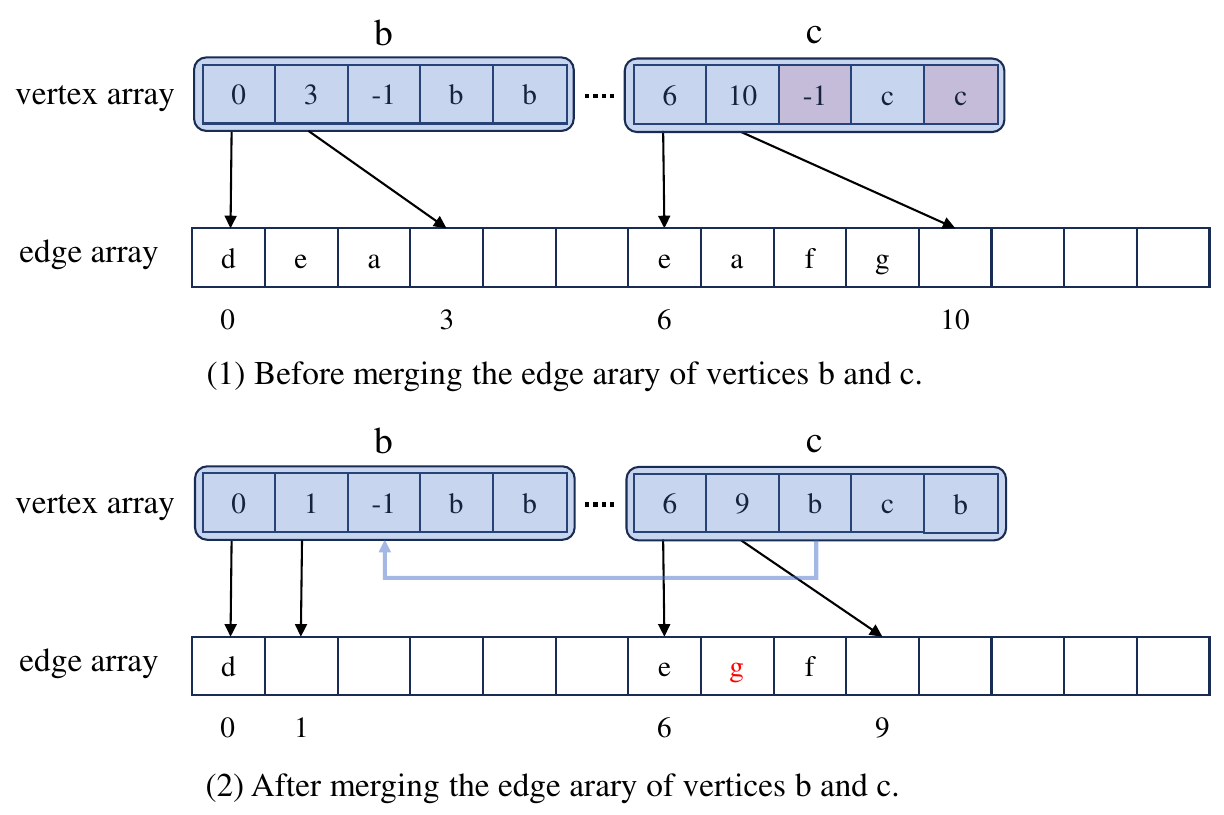}
    \caption{An example of connecting the edge tables of the boundary vertices, corresponding to the vertex merging operation in Figure \ref{fig:sigmod_kasi_rule}. Each vertex in the vertex array contains five pointers: $vtx\_ptr\_start$, $vtx\_ptr\_end$, $vtx
    \_link\_next$, $vtx\_link\_cur$, and $vtx\_link\_last$. The first two pointers are related to the positions of the elements in the edge array pointed to by the current vertex, while the last three pointers are associated with other edge tables connected to the current vertex. }    \label{fig:sigmod_ds_merge}
\end{figure}

We connect the edge tables by linking the boundary vertices. To minimize the cost of subsequent update operations, we remove vertices with low degrees. In the example shown in Figure \ref{fig:sigmod_kasi_rule}, vertex $b$ will be removed. 

As illustrated in Figure \ref{fig:sigmod_ds_merge}, we first use the value of $vtx\_link\_last$ to locate the last vertex connected to vertex $c$. Since vertex $c$ has not yet participated in any merge operations, the vertex we find is $c$ itself. Consequently, we update both $vtx\_link\_next$ and $vtx\_link\_last$ of vertex $c$ to point to $b$, indicating that the edge table of vertex $b$ has been merged into vertex $c$. During the traversal of the graph algorithm, $vtx\_link\_next$ is used to process the next edge list of the current vertex, while $vtx\_link\_last$ is specifically used in merge operations to quickly locate the last connected vertex of the current vertex.

After connecting the vertices, we remove duplicates and deleted elements from the edge list. For the edge list of vertex $c$, the last element is recorded at the position of vertex 
$a$ (which has already been removed), and the value of $vtx\_ptr\_end$ is decremented by one. The elements in the edge list of vertex $b$ are handled similarly.

It is evident that the time complexity of the connection operation can be maintained at $O(\sum_{\Tilde{v}\in \Gamma(\hat{v})} deg(\Tilde{v}))$. Compared to explicitly merging two edge tables in a dynamic array, this vertex-linking method significantly reduces the number of write operations.

\subsection{Updating the edge tables of the external vertices.}
\begin{figure}
\includegraphics[width=0.9\linewidth]{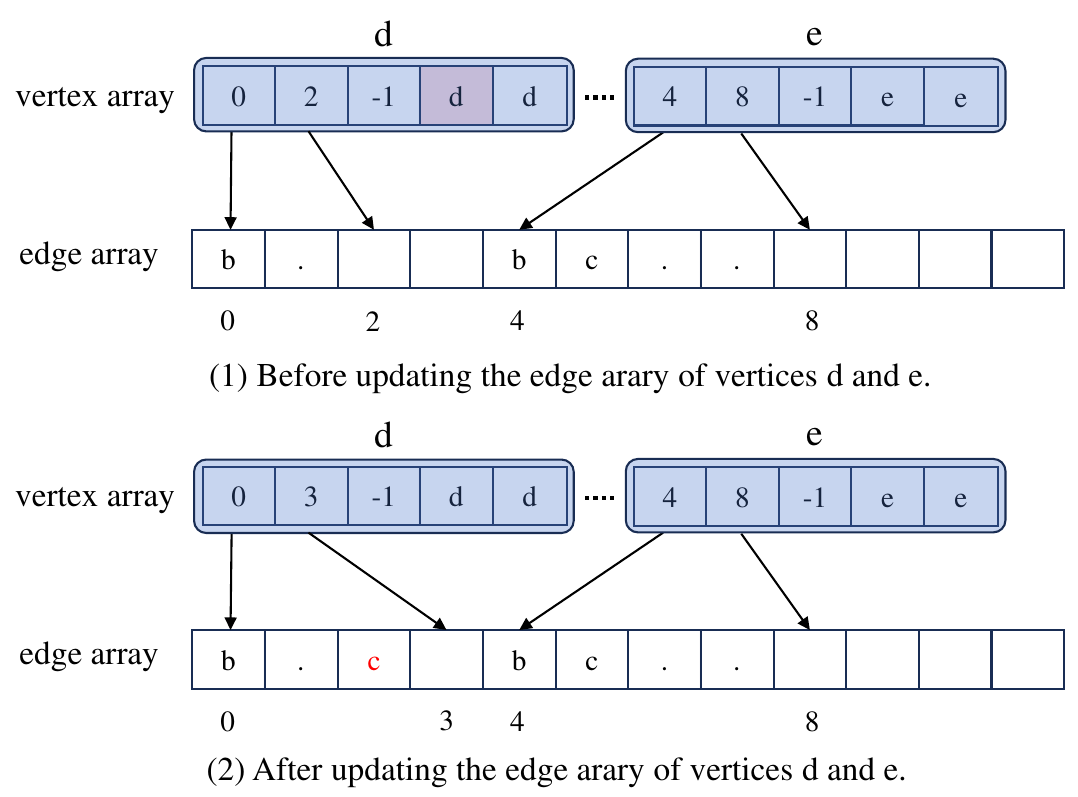}
\caption{An example of updating the edge tables for external vertices, corresponding to the vertex merging operation shown in Figure \ref{fig:sigmod_kasi_rule}. In the edge array, $.$s represent connections that are unaffected by the merge operation, corresponding to the dashed lines in Figure \ref{fig:sigmod_kasi_rule}.}
\label{fig:sigmod_ds_update}
\end{figure}

The update operation primarily targets the neighboring vertices of the removed boundary vertex, as they may need to establish a connection with the retained boundary vertex (prioritizing the removal of low-degree boundary vertices aims to minimize the time cost incurred by this process). For example, as shown in Figure \ref{fig:sigmod_kasi_rule}, vertices $d$ and $e$ are involved. Since vertex $e$ is already connected to $c$, no change in its connection status is necessary. We only need to consider how to establish the connection from vertex 
$d$ to vertex $c$.

It is easy to observe that if a external vertex requires a new edge to be constructed, it must already contain a removed boundary vertex in its edge table (this follows from the nature of the merge operation). For example, in Figure \ref{fig:sigmod_ds_update}, vertex $d$'s edge table contains vertex $b$, which is a removed vertex. We can record vertex $c$ in that position. However, in more general cases, locating the position of the removed boundary vertex might incur additional overhead due to accessing unrelated vertices. To ensure that this update operation can be performed in constant time, we designed a strategy to reduce the amortized cost of constructing each edge through batch processing.

As shown in Figure \ref{fig:sigmod_ds_update}, whenever a new edge needs to be constructed, we first use the value of $vtx\_link\_cur$ to locate the first connected vertex with an available gap (this gap could result from a prior removal operation or have been reserved during the construction of the edge table). Then, $vtx\_ptr\_end$ of the located vertex is used to find an insertable position. If no position is available, subsequent connected vertices will be accessed, and the value of $vtx\_link\_cur$ will be updated simultaneously. If none of the connected vertices' edge tables have available space, we will remove the deleted or duplicate elements from all edge tables. It's easy to understand that when the vertex currently being processed has an available gap, the time complexity for inserting an element is $O(1)$. The key issue lies in the time cost associated with the search operation when no gaps are available in the current vertex and the removal operation when none of the edge tables have available gaps. We analyze this by examining the amortized time complexity of each insertion operation.

Assume there are $x$ elements to be inserted into vertex $v$, with a total gap size of $y$ in the edge tables of vertex $v$, and $z$ edge tables connected to vertex $v$. It's easy to see that the edge tables of the vertex will be processed in at most $\frac{x}{y}$ rounds because each time the edge table is filled and a removal operation occurs, a gap equal to or larger than the previous one is created. For inserting these $x$ elements, the time complexity for finding the first gap in the edge table using the value of $vtx\_ptr\_end$ and inserting the element can be expressed as $x*O(1)$. The time complexity for finding the first edge table with a gap using $vtx\_link\_cur$ can be expressed as $\frac{x}{y}*z=\frac{x}{y}*O(deg(v))$ (where the number of connected edge tables is always less than the degree of $v$). Once the vertex's gaps are filled, the subsequent removal operation requires accessing the entire edge table, leading to a time complexity of $\frac{x}{y}*O(deg(v)+y)$.

Based on the reasoning above, the amortized time cost for each insertion can be expressed as follows.
\begin{equation}\label{equ:space}
\renewcommand{\arraystretch}{1.5}
\begin{tabular}{l}
$T_{amort.}=\lim_{x \to \infty}\frac{x*O(1)+\frac{x}{y}*O(deg(v))+\frac{x}{y}*O(deg(v)+y)}{x}$\\ 
$\qquad\quad\ =\lim_{x \to \infty}\frac{x*O(1)+\frac{x}{y}*O(deg(v)+y)}{x}$\\
$\qquad\quad\ =\lim_{x \to \infty} 
 O(1)+O(\frac{deg(v)}{y})$ 
\end{tabular}
\end{equation}

From Equation \ref{equ:space}, we can observe that when the gap size is on the same order of magnitude as $deg(v)$, the amortized cost for updating each external vertices can be maintained at $O(1)$. In this case, the time cost incurred by the update operation will not affect the overall time complexity of the algorithm. The constant time complexity is essentially achieved by reducing the frequency of search operations. For each inserted boundary vertex, the corresponding deleted boundary vertex is not immediately searched for removal. Instead, it is removed during a later update operation applied to the entire edge table. This approach reduces the average cost of removing each deleted boundary vertex from the edge table. In our practical experiments, since the first data reduction rule inherently create many gaps, we did not increase the size of the edge table. However, if a strict time complexity guarantee is required, this can be achieved by doubling the size of the edge table.

\section{Evaluation}
\label{sec:exp}

\noindent\textbf{Dataset.} We conducted experimental evaluations on twenty-six real-life graphs and eight synthetic graphs. The real-life graphs are derived from matrices related to graphs and networks~\cite{snapnets} in the University of Florida Sparse Matrix Collection~\cite{Davis2011TheUO}, while the synthetic graphs were constructed based on the specific instances proposed by Kaya et al\cite{Kaya2020KarpSipserBK}.

The real-life graphs includes all matrices in the SNAP category~\cite{snapnets} with vertex counts ranging from one million to ten million and edge counts less than two hundred million, totaling twelve matrices. These matrices have been widely used in previous studies\cite{Duff2011DesignIA,Azad2016DistributedMemoryAF,Kaya2020KarpSipserBK} to evaluate the performance of bipartite graph matching. Directed and undirected graphs can be used to construct bipartite graphs because each vertex in a directed graph can naturally be split into two vertices in a bipartite graph based on its in-degree and out-degree\cite{Liu2011ControllabilityOC,Vazifeh2018AddressingTM}, while undirected graphs can be treated as directed graphs with bidirectional edges. Detailed information about these matrices is provided in Table \ref{tab:dateset}. 

Each synthetic graph is composed of 64 special instances proposed by Kaya et al.~\cite{Kaya2020KarpSipserBK}, which can cause the $KaSi$ algorithm to exhibit its worst-case time complexity. Additionally, for n-2 of the mergeable vertices in each special instance, we added an extra edge to reduce the number of explicit mergeable vertices. We constructed a total of eight such synthetic graphs, with the number of vertices growing exponentially from $2^{15}$ to $2^{22}$. To test the algorithm's robustness, we followed previous research\cite{Duff2011DesignIA, Kaya2020KarpSipserBK} by preprocessing the real-life matrices inputs with random permutations. All matrices were provided to the algorithm in CSR format. In the subsequent experiments, each instance was run five times, and the average value was taken. All runtime measurements are reported in seconds.

\begin{table}[]
\centering
\caption{Dateset}
\begin{tabular}{c|c|c|c} 
\hline
name & n & m & kind \\ \hline
as-Skitter\cite{Leskovec2005GraphsOT} & 1,696,415 & 22,190,596 & Undirected  \\ \hline
cit-Patents\cite{Leskovec2005GraphsOT}  &3,774,768 & 16,518,948 & Directed   \\ \hline
com-LiveJournal\cite{Yang2012DefiningAE} &3,997,962	 &69,362,378 & Undirected  \\ \hline
com-Youtube\cite{Yang2012DefiningAE} & 1,134,890 & 5,975,248	& Undirected  \\ \hline
ljournal-2008\cite{Boldi2004TheWF,Boldi2010LayeredLP} & 5,363,260 & 79,023,142 & Directed  \\ \hline
roadNet-CA\cite{Leskovec2008CommunitySI} & 1,971,281	 & 5,533,214 & Undirected  \\\hline
roadNet-PA\cite{Leskovec2008CommunitySI} & 1,090,920	 & 3,083,796 & Undirected \\ \hline
roadNet-TX\cite{Leskovec2008CommunitySI} & 1,393,383& 3,843,320	& Undirected \\ \hline
soc-LiveJournal1\cite{Leskovec2008CommunitySI,Backstrom2006GroupFI} & 4,847,571 & 68,993,773 & Directed \\ \hline
soc-Pokec\cite{Takac2012DATAAI} &1,632,803  &  30,622,564	& Directed  \\ \hline
wiki-Talk\cite{Leskovec2010SignedNI,Leskovec2010PredictingPA} & 2,394,385 &  5,021,410& Directed \\ \hline
wiki-topcats\cite{Yin2017LocalHG,Klymko2014UsingTT} & 1,791,489& 28,511,807 &  Directed \\ \hline
\end{tabular}    
\label{tab:dateset}
\end{table}

\noindent\textbf{Evaluated algorithms and implementation.} In the first part, we evaluated algorithms implementing Karp and Sipser's data reduction rules. We first verified the effectiveness of our designed strategies. Then we compared $MVM$ (including all three strategies we proposed) with $KaSi\_cache$, $KaSi\_comp$ (single-threaded version), $HKaSi$, and $TKaSi$. Among them, $KaSi\_cache$\cite{Kaya2020KarpSipserBK} and $KaSi\_comp$\cite{Langguth2021SharedmemoryIO} are the most advanced implementations currently available, while $HKaSi$ and $TKaSi$ are two theoretical algorithms we realized using hash tables and red-black trees (with the same search efficiency as a binary search tree) as storage structures. In the second part, we evaluated the acceleration effect of the kernelization algorithm for obtaining maximum matchings in bipartite graphs. We compared the kernelization algorithm with state-of-the-art maximal matching algorithms, which were also proposed to accelerate maximum matching computation. We compared $MVM$ with $TruncRW$\cite{Panagiotas2020EngineeringFA} and $MatchBG$\cite{GuangWu24}. The former obtains a maximal matching through a truncated random walk method based on doubly stochastic matrices, while the latter uses Karp and Sipser's first rule and the crown rule to obtain a high-quality maximal matching. The exact algorithm we used is $PFP$\cite{Duff2011DesignIA}, which has been recognized as one of the most effective algorithms in recent experimental studies\cite{Kaya2020KarpSipserBK, Panagiotas2020EngineeringFA}. All algorithms were either implemented using the source code provided in the paper or developed in C/C++ following a similar style. The code was executed on a machine running Ubuntu 20.04.6 LTS, featuring an i9-14900 CPU and 32GB of RAM.

\subsection{Comparison with KaSi’s variants}
In this section, we will compare $MVM$ against the state-of-the-art implementation of the $KaSi$ algorithm and its variants on real-life and synthetic graphs.

\begin{figure}
    \centering
\includegraphics[width=\linewidth]{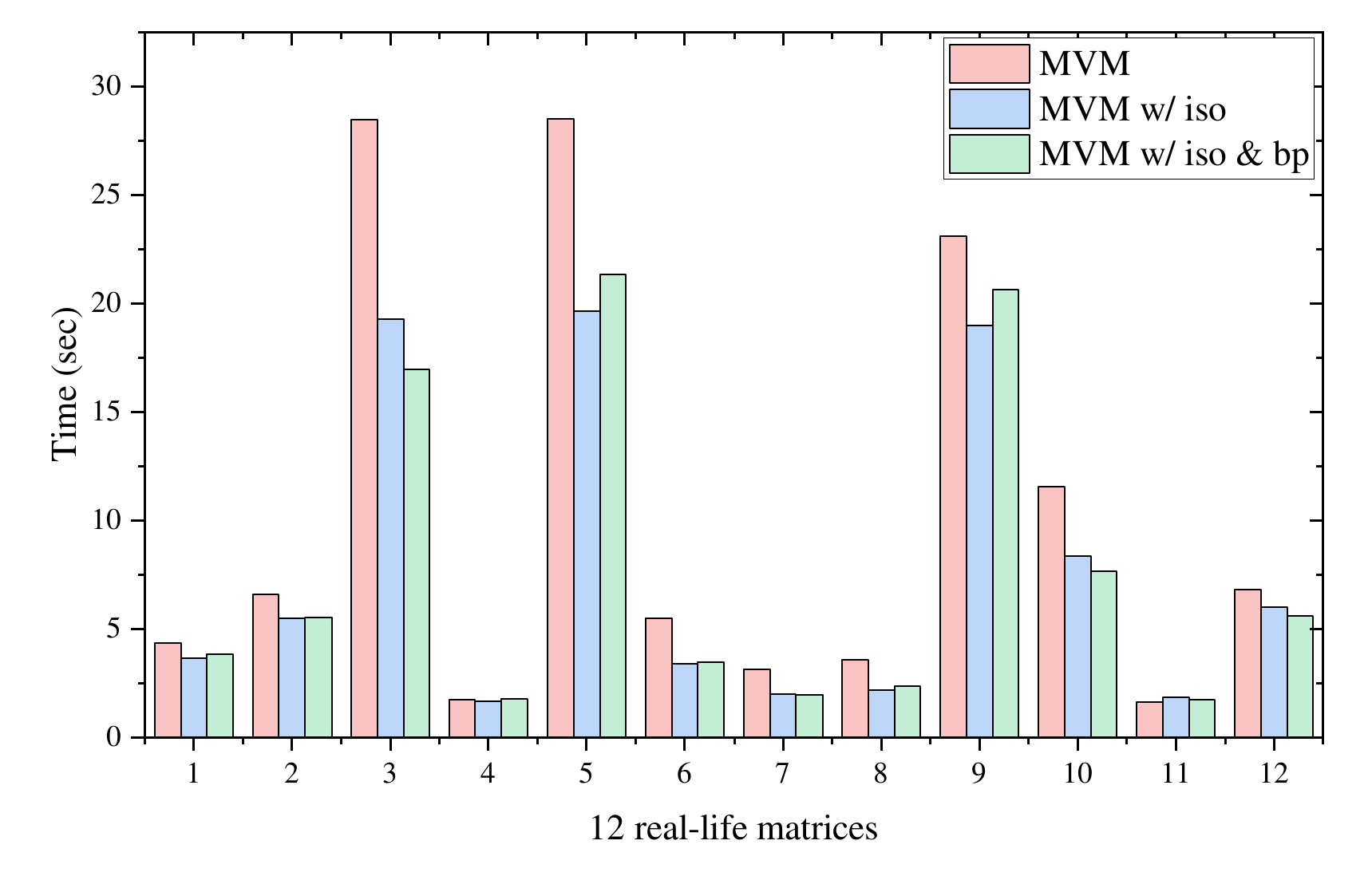}
    \caption{The runtime of MVM with different optimization strategies.}
    \label{fig:mvm_ablation}
\end{figure}

\noindent\textbf{Ablation study.} As shown in the Figure \ref{fig:mvm_ablation}, we demonstrated the impact of different optimization strategies. It's easy to observe that the optimization effect brought by indirect set operations (iso) is the most significant, indicating that using the iso strategy can indeed avoid set operations between current mergeable vertex sets and irrelevant vertices. The $MVM$ using the balanced processing (bp) strategy performs slightly worse than the $MVM$ using the greedy strategy on some instances. This is because the balanced processing strategy was proposed to avoid the worst-case time complexity in theory. In real-life graphs, such high-degree boundary vertices might be rare, and the balanced processing strategy slightly increases the number of merge operations compared to the greedy strategy, thus causing a minor increase in time consumption on real-life graphs. We believe it's worthwhile to accept this slight increase in time consumption to achieve a lower upper bound on time complexity. In subsequent experiments, $MVM$ refers to the kernelization algorithm that incorporates all three optimization strategies.

\begin{figure}
    \centering
    \includegraphics[width=\linewidth]{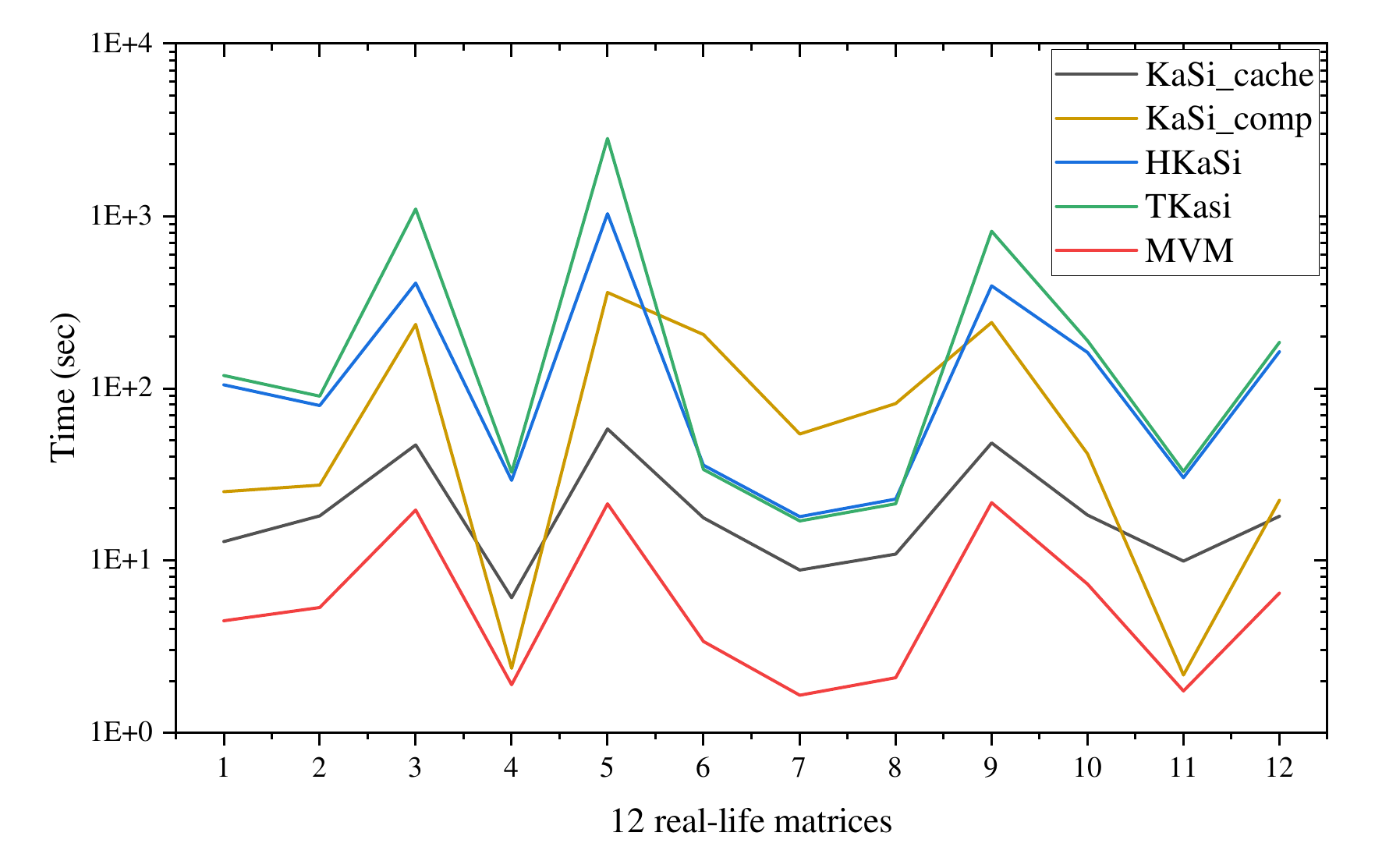}
    \caption{The runtime of MVM and the variants of KaSi algorithm.}
\label{fig:sigmod_all_kasi_reallife2}
\end{figure}
\noindent\textbf{On real-life graphs.} As shown in Figure \ref{fig:sigmod_all_kasi_reallife2}, we demonstrate the runtime of $MVM$ and the variants of $KaSi$. By analyzing the experimental data, we can draw the following conclusions.

1) $HKaSi$ and $TKaSi$ perform the worst on real-life graphs, demonstrating the drawbacks of non-sequential storage structures in graph algorithms. Hash tables and red-black trees are inefficient for the frequent neighborhood searches required. On average, their runtimes are 25 and 56 times longer than $MVM$, respectively, highlighting the limitations of relying on specialized storage structures to reduce time complexity.

2) $KaSi\_cache$ and $KaSi\_comp$ perform better than $HKaSi$ and $TKaSi$. However, algorithms without theoretical guarantees may lack robustness when facing extreme cases. Subsequent experiments show that when handling special instances demonstrating their time complexity, $KaSi\_cache$ is nearly 4,340 times slower than $MVM$ when the number of vertices reaches $2^{22}$. As the number of vertices increases, the runtime gap will further widen. $KaSi\_comp$, in addition to lacking a time complexity guarantee, performs worse than $KaSi\_cache$ due to its strategy of only expanding explicit mergeable vertices, whereas $KaSi\_cache$ benefits from directly caching high-degree vertices (which can be used for both explicit and implicit mergeable vertices).

3) $MVM$ performs the best on real-life graphs. Compared to $KaSi\_comp$, $MVM$ leverages a multi-vertex merging strategy and an indirect set operation approach to handle both explicit and implicit mergeable vertices, thereby reducing the frequency of storage structure modifications. Compared to $KaSi\_cache$, $MVM$ employs a more efficient data structure for read and write operation, significantly lowering the time cost of each merge operation. As a result, $MVM$ achieves the lowest runtime across all tested graphs.

\begin{figure}
    \centering
    \includegraphics[width=\linewidth]{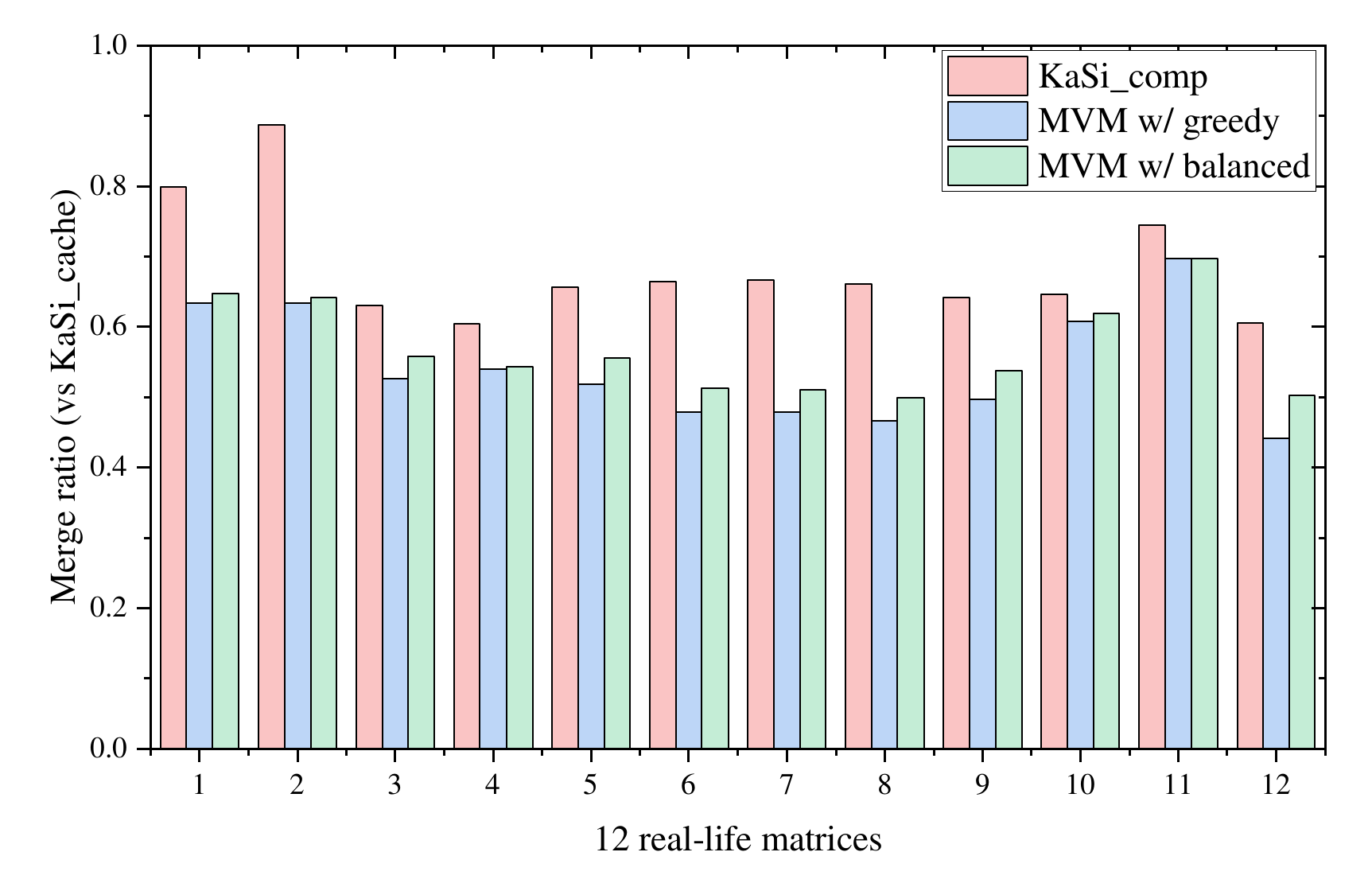}
    \caption{The merge ratio on real-life Graphs.}
    \label{fig:mergetimes}
\end{figure}

\noindent\textbf{The ratio of merge operations.} As shown in Figure \ref{fig:mergetimes}, we illustrate the ratio of merge operations performed by our algorithm compared to those by $KaSi$ and $KaSi\_comp$. It is evident that on most graphs, employing our proposed multi-vertex merge strategy can reduce the number of merge operations by nearly half, elucidating why our algorithm incurs lower time consumption. Although the merge ratio of $MVM$ w/ balanced is slightly higher than that of $MVM$ w/ greedy, Figure 4 shows that $MVM$ w / balanced still performs comparably to $MVM$ w/ greedy on real-life graphs. Therefore, we believe that this trade-off—slightly increasing the number of merge operations to achieve a lower upper bound on time complexity for sparse graphs—is worthwhile.

\begin{figure}
    \centering
    \includegraphics[width=0.9\linewidth]{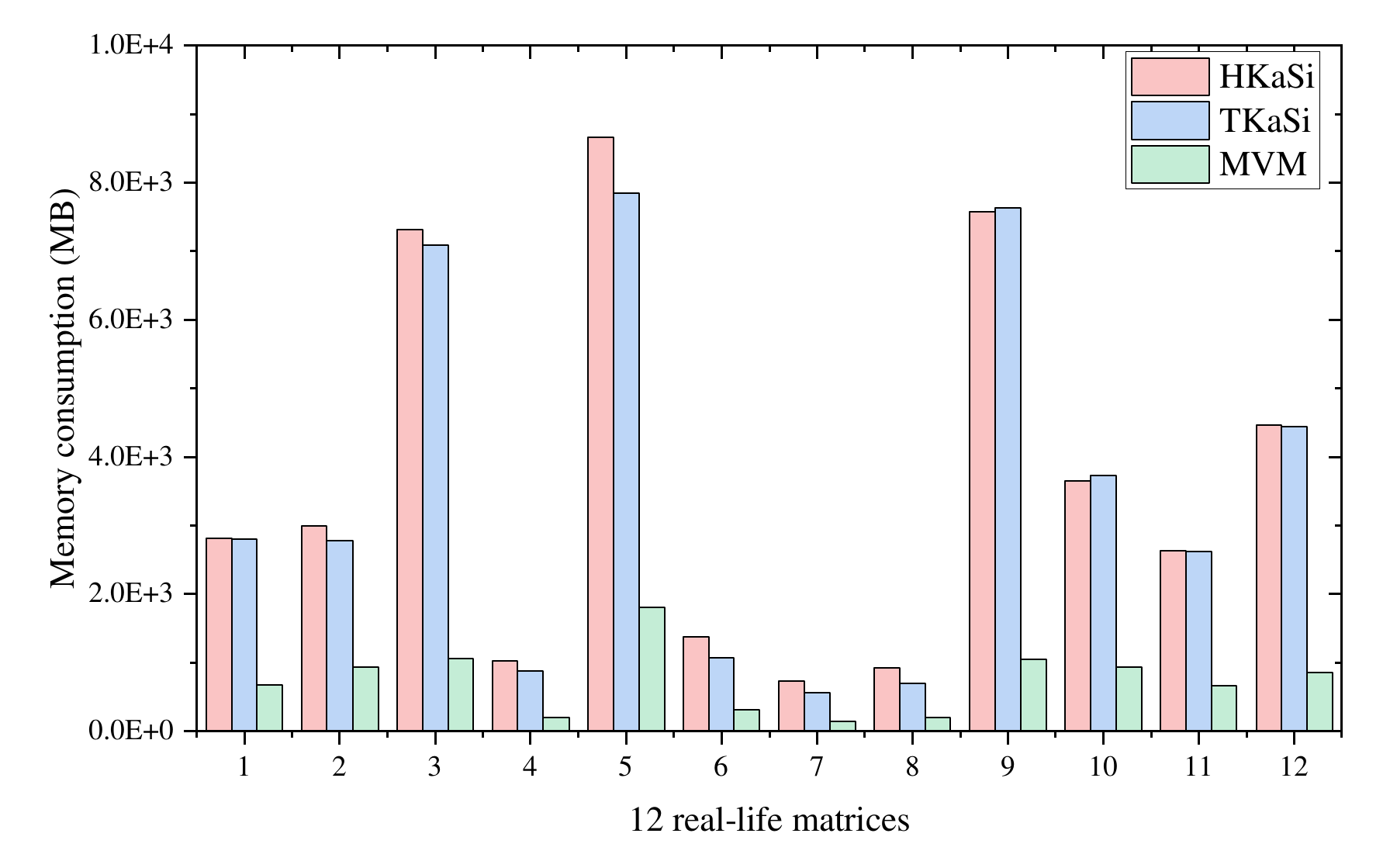}
    \caption{The memory consumption on real-life graphs.}
    \label{fig:sigmod_memory_comsuption}
\end{figure}

\noindent\textbf{The memory consumption.} As shown in Figure \ref{fig:sigmod_memory_comsuption}, we present the memory consumption of these kernelization algorithms. Since algorithms that use sequential storage structures (with O(n) search efficiency) have similar memory usage, we only compare $MVM$ with the two theoretical algorithms. It can be observed that $HKaSi$ and $TKaSi$, which use hash tables and red-black trees as storage structures, exhibit significantly higher memory consumption. In some graphs, their memory usage is nearly ten times that of $MVM$. Storage structures like CSR can achieve better traversal and space efficiency through compact data storage. This is why we aim to design algorithms with lower time complexity for data structures that maintain $O(n)$ search efficiency.

\begin{figure}
    \centering
    \includegraphics[width=0.85\linewidth]{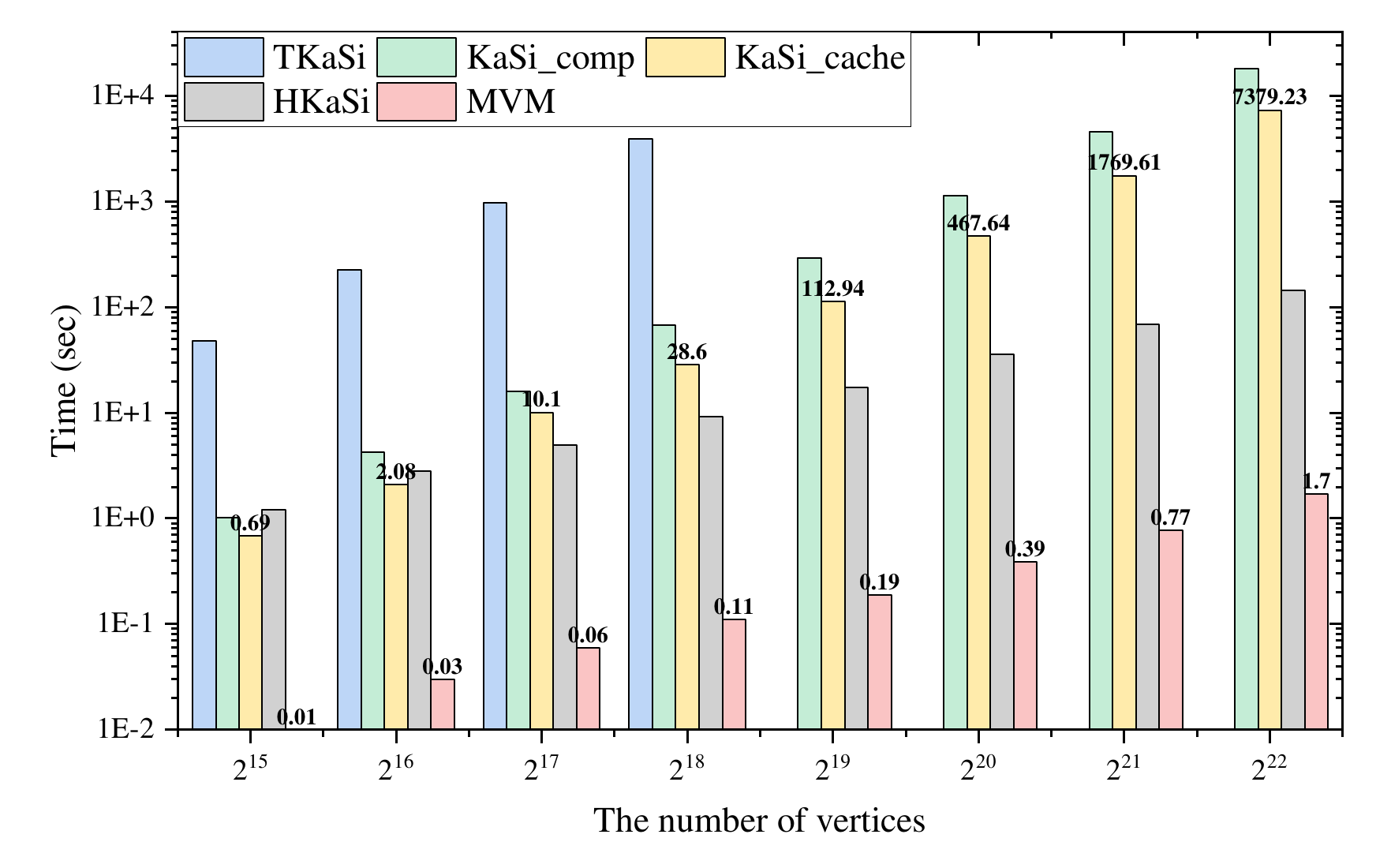}
    \caption{The runtime on the worst-case instances.}
    \label{fig:sigmod_worstcase}
\end{figure}

\noindent\textbf{On the worst-case instances.} As shown in Figure \ref{fig:sigmod_worstcase}, we present the runtime of these kernelization algorithms on special instances that exhibit their worst-case time complexity. Due to the excessively high runtime of $TKaSi$ on graphs with more than $2^{18}$ vertices, its measurement results are not included in the subsequent examples.

The runtime growth rate of $KaSi\_cache$ and $KaSi\_comp$ is nearly quadratic with respect to the increase in data size, which aligns with their theoretical worst-case time complexity. While $KaSi\_cache$ alleviates some of the impact of high-degree boundary vertices by caching them, once the number of high-degree boundary vertices exceeds the cache size (which we set to 10 edge tables), no caching strategy can prevent performance degradation. $KaSi\_comp$, on the other hand, suffers from repeatedly processing the same boundary vertices because it fails to recognize the implicit mergeable vertices, leading to the poor performance.

As the number of vertices increases, the runtime growth rate of $TKaSi$ gradually approaches linearity. However, due to its high initial runtime, $TKaSi$ performs the worst on these instances. Both theoretically and practically, $TKaSi$ is outperformed by $HKaSi$. $HKaSi$, which uses hash tables as its storage structure, significantly mitigates the worst-case scenarios. However, since hash tables are not well-suited for sequential access, its runtime remains one to two orders of magnitude higher than that of $MVM$. Due to its theoretical time complexity guarantee and efficient implementation, $MVM$ demonstrates excellent scalability and robustness on these worst-case instances, and it performs best across all test cases, whether on real-life graphs or synthetic graphs.

\subsection{Comparison with maximal matching algorithms}

Similarly to the kernelization algorithm, maximal matching algorithms are also employed to expedite the process of obtaining the maximum matching. In this section, we will compare the kernelization algorithm with the state-of-the-art maximal matching algorithms. Since the kernel graphs produced by the kernelization algorithms in the previous section have equal sizes, we will solely use $MVM$ to compare against these maximal matching algorithms.

\begin{figure}
    \centering
    \includegraphics[width=0.9\linewidth]{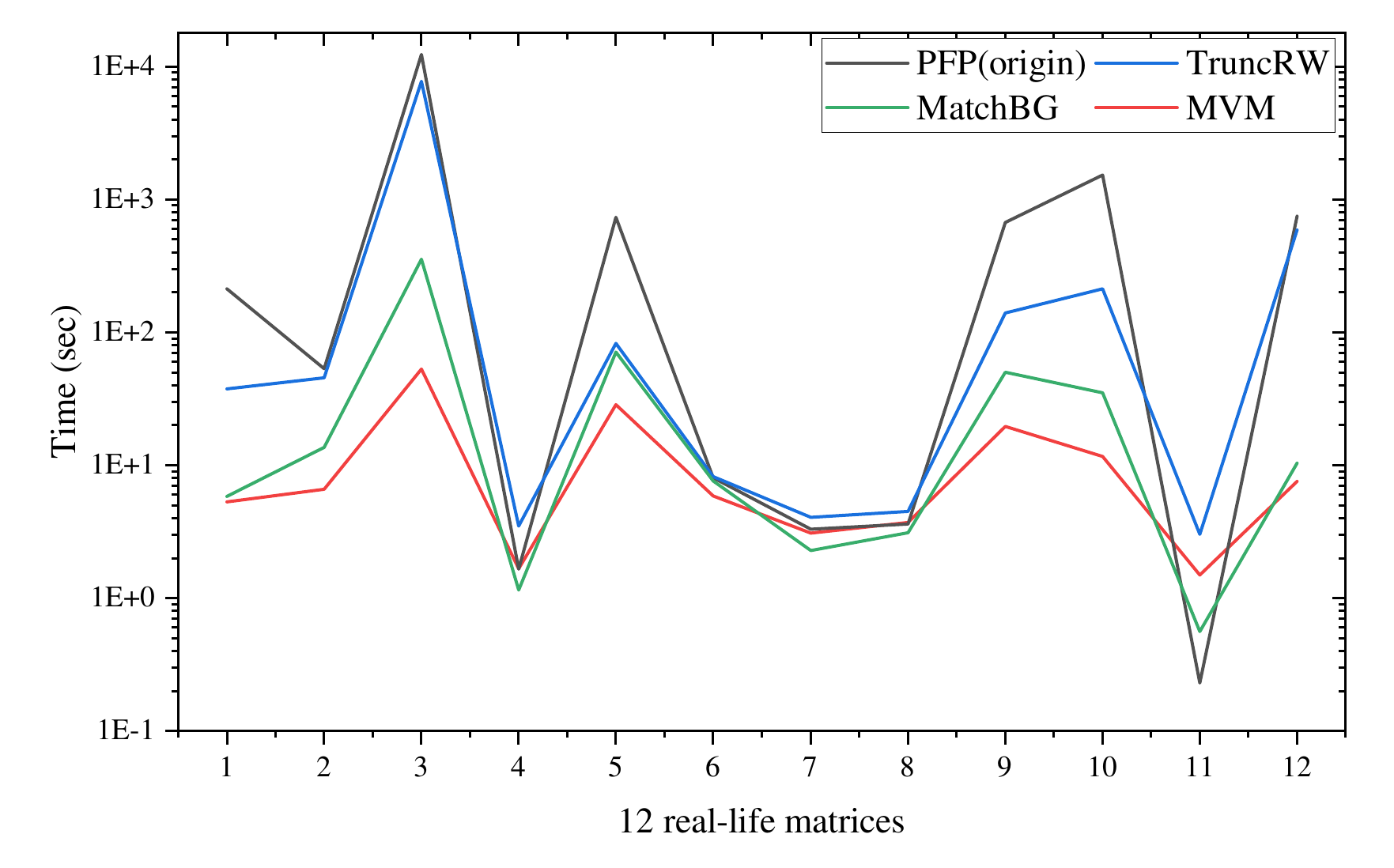}
    \caption{The overall runtime for obtaining the maximum matching on real-life graphs, where $origin$ indicates the absence of acceleration methods.}
    \label{fig:sigmod_vs_maximal}
\end{figure}

\noindent\textbf{The overall performance.} As shown in Figure \ref{fig:sigmod_vs_maximal}, we can observe that in most cases, $MVM$ achieves the best performance. On graphs where the exact algorithm ($PFP$) can quickly find the maximum matching, the total runtime primarily depends on the execution time of the acceleration method. Due to our efficient implementation of $MVM$, the kernelization method remains competitive with the state-of-the-art maximal matching algorithms. On graphs where the exact algorithm performs poorly, $MVM$ achieves significant acceleration compared to the maximal matching algorithms, as the kernelization algorithm effectively reduces the size of the input data, thereby lowering the cost of obtaining an exact solution. Overall, on nearly all real-life graphs, when $MVM$ is used as an acceleration method, the total runtime remains very low, typically under 64 seconds. This highlights the advantage of using kernelization algorithms to accelerate the process of obtaining the maximum matching.

\begin{figure}
    \centering
    \includegraphics[width=0.9\linewidth]{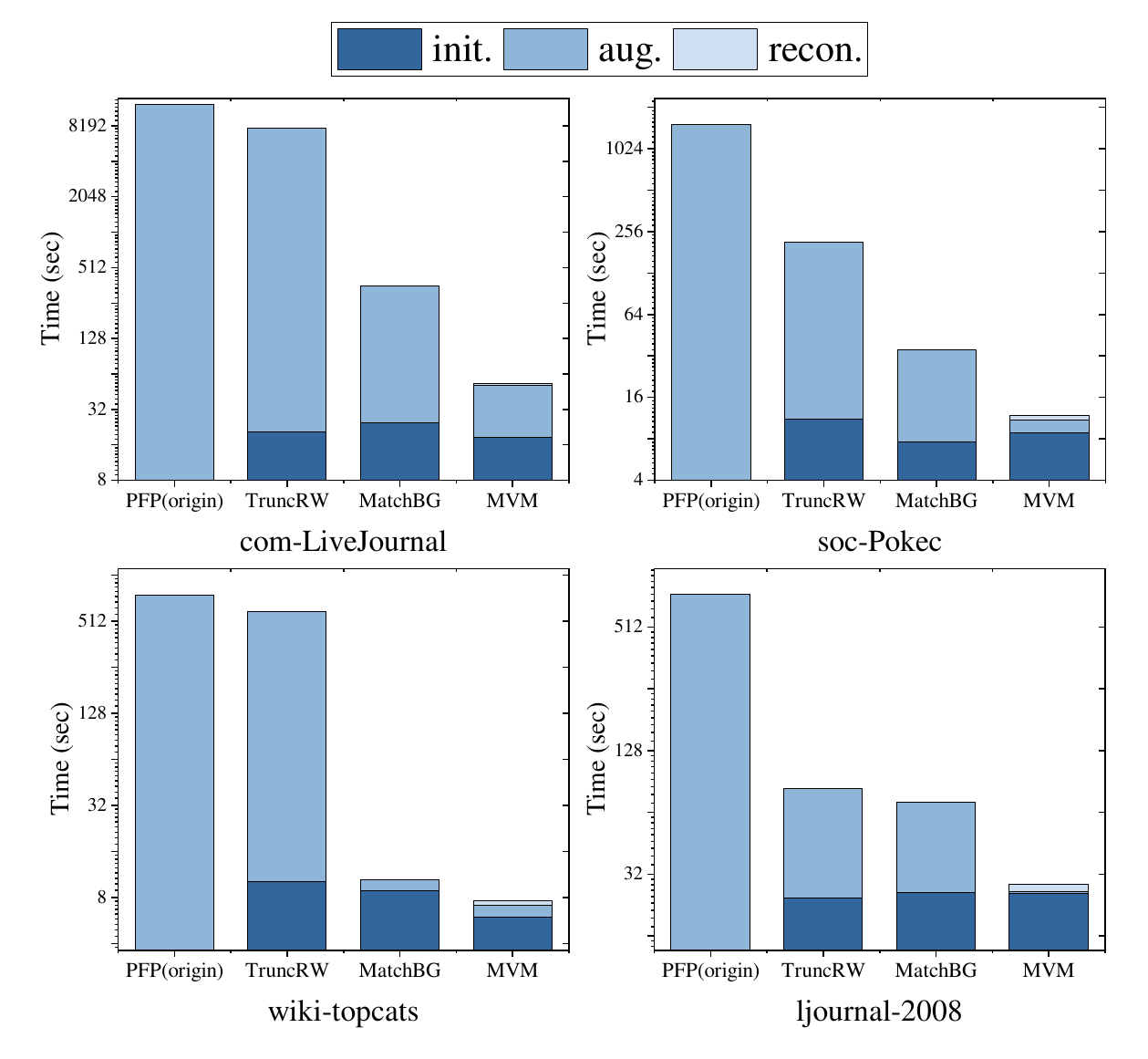}
    \caption{The detailed runtime of the four worst-performing graphs when using the exact algorithm to obtain the maximum matching. $init.$ represents the time cost for maximal matching algorithms and kernelization algorithm, $aug.$ represents the time cost for the exact algorithm, $recon.$ represents the time cost for reconstructing maximum matching (applicable only to $MVM$).
    }
\label{fig:sigmod_reallife_detail}
\end{figure}

\noindent\textbf{Detailed study.} To conduct a detailed analysis, we present the four worst-performing matrices when using the exact algorithm to obtain the maximum matching. As shown in Figure \ref{fig:sigmod_reallife_detail}, we can get the following observations.

1) On these graphs where obtaining the maximum matching is challenging, the overall running time is predominantly determined by the time consumed by the exact algorithm during the augmentation step, while the time expended in other steps can be disregarded. Therefore, it is justified to utilize a cheap algorithm to accelerate the process of the exact algorithm.

2) The acceleration provided by the random selection-based $TruncRW$ is limited. As the size of the matching set increases, the cost of adding matching edges also rises. $TruncRW$ essentially captures the matching edges that are easily obtained. Moreover, since $TruncRW$ cannot reduce the search space, subsequent exact algorithms may need to construct long augmenting paths when adding matching edges, which can lead to significant time consumption    

3) The acceleration effect of the quality-focused maximal matching algorithm $MatchBG$ is weaker than that of $MVM$. This is because $MatchBG$ only applies a subset of Karp and Sipser's data reduction rules, specifically those that do not require merging the edge table to modify the storage structure. Furthermore, $MatchBG$ has a time complexity of $O(n\sqrt{n}*\mathcal{D})$, which is approximately $O(m\sqrt{n})$, making it less competitive compared to $MVM$’s $O(m \log n)$. Consequently, both in theory and practice, $MVM$ consistently outperforms $MatchBG$.

4) The kernelization algorithm delivers significant acceleration. On com-LiveJournal, applying $MVM$ as a preprocessing step before the exact algorithm results in an approximately 230-fold speedup. For other graphs where obtaining the maximum matching is particularly challenging, the kernelization method can reduce overall runtime to mere tens of seconds or even less. While maximal matching algorithms are primarily advantageous due to their low computational overhead, our highly optimized $MVM$ implementation effectively eliminates this edge. Moreover, our kernelization algorithm provides more consistent and reliable acceleration by substantially reducing the size of the input problem.

\begin{figure}
    \centering
    \includegraphics[width=0.85\linewidth]{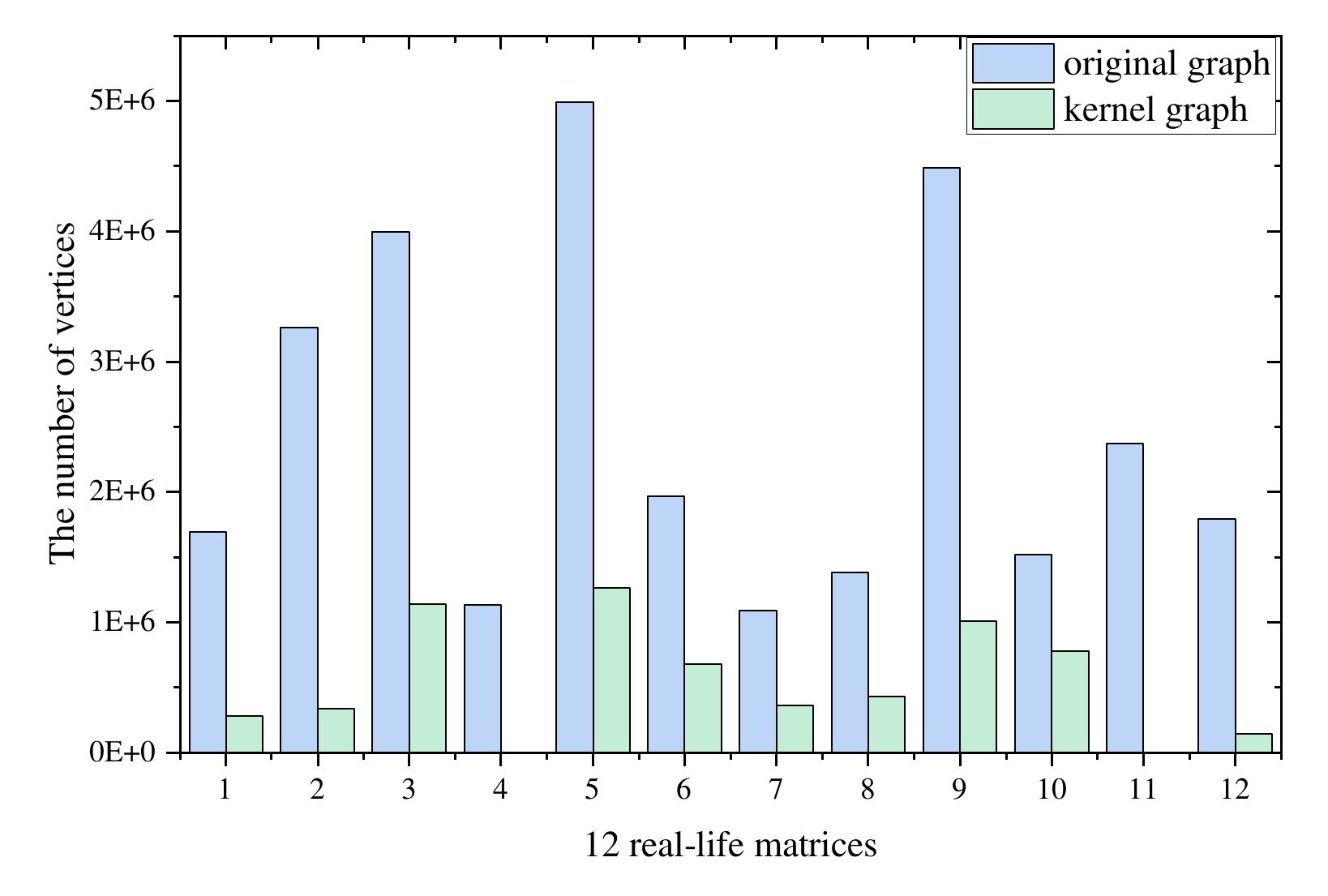}
    \caption{The kernelization quality on real-life graphs.}
\label{fig:sigmod_kernel_quality}
\end{figure}

\noindent\textbf{Kernelization quality.} To illustrate why kernelization algorithms achieve better acceleration compared to maximal matching algorithms, we present the kernelization quality of $MVM$. As shown in Figure \ref{fig:sigmod_kernel_quality}, kernelization significantly reduces the number of vertices in most graphs. In contrast, applying maximal matching algorithms does not alter the graph's size. Beyond reducing data size, kernelization also increases the graph's density. Poloczek et al.\cite{Poloczek2012RandomizedGA} demonstrated that obtaining the maximum matching is easier in relatively denser graphs, which explains why kernelization algorithms can achieve such acceleration.

In practice, maximal matching algorithms can be further applied to the kernel graphs obtained through kernelization to achieve additional acceleration. Thanks to our highly optimized implementation of the $MVM$ algorithm, $MVM$ can act as an assistant to maximal matching algorithms rather than a competitor. However, for fairness in this study, we did not apply maximal matching algorithms to the kernel graphs.

\section{Conclusion}
We investigate the implementation of Karp and Sipser's data reduction rules on bipartite graphs, which are proposed to accelerate the acquisition of the maximum matching. We introduced the algorithm $MVM$, which is the first to apply these rules exhaustively with nearly linear time complexity, while previous implementations had a time complexity of $O(n^2)$. Additionally, we designed a storage structure that efficiently supports vertex merging operations while ensuring data locality, thereby improving traversal efficiency. Finally, through extensive experiments on real-life and synthetic graphs, we demonstrated the superiority of our proposed algorithms.

\newpage
\bibliographystyle{ieeetr}
\bibliography{sample-base}

\end{document}